\documentclass[twocolumn]{aastex63}
\shortauthors{Qin et al.}

\begin{document}

\title{On the Angular Momentum Transport Efficiency within the Star Constrained from Gravitational-wave Observations}

\email{yingqin2013@hotmail.com}
\email{wangyz@pmo.ac.cn}
\email{wudonghong@ahnu.edu.cn}

\author[0000-0002-2956-8367]{Ying\,Qin}
\affiliation{Department of Physics, Anhui Normal University, Wuhu, Anhui, 241000, PR, China}
\affiliation{Guangxi Key Laboratory for Relativistic Astrophysics, Nanning 530004, China}

\author[0000-0002-2956-8367]{Yuan-Zhu\, Wang}
\affiliation{Key Laboratory of Dark Matter and Space Astronomy, Purple Mountain Observatory, Chinese Academy of Sciences, Nanjing, 210033, People's Republic of China}
\author[0000-0001-9424-3721]{Dong-Hong\, Wu}
\affiliation{Department of Physics, Anhui Normal University, Wuhu, Anhui, 241000, PR, China}
\affiliation{School of Astronomy and Space Science and Key Laboratory of Modern Astronomy and Astrophysics in Ministry of Education, Nanjing University, Nanjing 210093, China}

\author{Georges\, Meynet}
\affiliation{Geneva Observatory, Geneva University, CH-1290 Sauverny, Switzerland}

\author{Hanfeng\, Song}
\affiliation{College of Physics, Guizhou University, Guiyang city, Guizhou Province, 550025, P.R. China}


\begin{abstract}
The LIGO Scientific Collaboration and Virgo Collaboration (LVC) have recently reported in GWTC-2.1 eight additional candidate events with a probability of astrophysical origin greater than 0.5 in the LVC deeper search on O3a running. In GWTC-2.1, the majority of the effective inspiral spins ($\chi_{\rm eff}$) show magnitudes consistent with zero, while two (GW190403$_{-}$051519 and GW190805$_{-}$211137) of the eight new events have $\chi_{\rm eff}$ $> 0$ (at 90\% credibility). We note that GW190403$_{-}$051519 was reported with $\chi_{\rm eff}$ = $0.70^{+0.15}_{-0.27}$ and mass ratio $q$ = $0.25^{+0.54}_{-0.11}$, respectively. Assuming a uniform prior probability between 0 and 1 for each black hole's dimensionless spin magnitude, GW190403$_{-}$051519 was reported with the dimensionless spin of the more massive black hole, $\chi_1$ = $0.92^{+0.07}_{-0.22}$. This is the fastest black hole ever measured in all current gravitational-wave events. If the immediate progenitor of GW190403$_{-}$051519 is a close binary system composed of a black hole and a helium star, which can be the natural outcome of the classical isolated binary evolution through the common envelope phase, this extremely high spin challenges, at least in that case, the existence of efficient angular momentum transport mechanism between the stellar core and the radiative envelope of massive stars, as for instance predicted by the Tayler-Spruit dynamo \citep{2002A&A...381..923S} or its revised version by \cite{2019MNRAS.485.3661F}.
\end{abstract}


\section{Introduction}
The detection of the first gravitational wave (GW) event from the coalescence of two black holes (BHs), GW150914 \citep{2016PhRvL.116f1102A}, has opened a new window to directly study BHs. By the end of the first half of the third observing run (O3a), the Advanced LIGO \citep{2015CQGra..32g4001L} Scientific Collaboration and Advanced Virgo \citep{2015CQGra..32b4001A} Collaboration (LVC) had reported on 46 binary black hole (BBH) events and one event (GW190814) with an unidentified lighter component in the second Gravitational-Wave Transient Catalog (GWTC-2) \citep{2021PhRvX..11b1053A}. Recently eight additional candidate events that passed a false alarm rate threshold of two per day during O3a running have been reported with a probability of astrophysical origin greater than 0.5 \citep{2021arXiv210801045T}. The majority of the effective spins have magnitudes similar to the events reported in GWTC-2 \citep{2021PhRvX..11b1053A}, being consistent with zero. 

The effective inspiral spin $\chi_{\rm eff}$ that can be directly constrained by the gravitational wave signal, is defined as
\begin{equation}\label{xeff}
\end{equation}
where $M_1$ and $M_2$ are the primary and secondary masses, $\vec{\chi_1}$ and $\vec{\chi_2}$ are the dimensionless spin \footnote{The astrophysical BH can be fully described by its mass $M$ and angular momentum $\vec{J}$. The BH spin $\vec{\chi} = c \vec{J}/G M^2$, where $c$ is the speed of light in vacuum, $G$ the gravitational constant.} parameters, and $\hat{L}_N$ is the unit vector along the orbital angular momentum.

Of the eight confident candidates, GW190403$_{-}$051519 and GW190805$_{-}$211137, were reported with $\chi_{\rm eff}$ $>$ 0 at the 90\% credible level. Interestingly, GW190403$_{-}$051519 shows a very unequal mass ratio ($q$ = $0.25^{+0.54}_{-0.11}$), and the highest effective spin ($\chi_{\rm eff}$ = $0.70^{+0.15}_{-0.27}$) ever reported by LIGO/Virgo. For binaries with very unequal masses, the spin magnitude of the more massive object dominates over the secondary on the effective spin, which thus permits stringent constraints on the primary spin magnitude. For GW190403$_{-}$051519, we have thus a very constrained value for $\chi_1$ of $0.92^{+0.07}_{-0.22}$.

More and more binary BH formation channels have been proposed since the discovery of GW150914 \citep{2016ApJ...818L..22A} and the effective spin is considered as an indicator of the formation path. In the canonical isolated binary evolution scenario \citep{1993MNRAS.260..675T,2016Natur.534..512B}, the more massive star (primary star) initially evolves in a wide orbit to become a red supergiant star. It then becomes a helium star after losing its outer layers due to the stellar winds and/or mass transfer to its companion (secondary star) via the first Lagrangian point ($L_1$). The helium star soon directly collapses to form a BH (first-born BH). The first-born BH accretes negligible material during the so-called Common Envelope (CE) \citep{2013A&ARv..21...59I} phase, and thus the change of the spin and mass of the first-born BH is negligible. Therefore the first-born BH spin is natal and is inherited from the angular momentum content of its progenitor. The progenitor cannot obtain any angular momentum due to negligible tides and winds mass-loss, which further signifies that the rotation rate is exclusively determined by the angular momentum of the progenitor itself. We note that a recent population synthesis study \citep{2021ApJ...921L...2O} with efficient angular momentum transport assumption can still reproduce highly spinning more massive BHs in the isolated binary evolution channel. In their proposed channels, the fast-spinning more massive BHs in BBH system can be formed either from the stable mass transfer (see Fig. 1), or from two equal-mass helium stars spun up by the tides just after the CE phase (see Fig. 2). However, both channels are different from the classical isolated binary evolution scenario we study in this work, which is used for testing the formation of GW190403$_{-}$051519.

The first-born BH will have negligible spin \citep[e.g., see Fig.1 in][]{2018A&A...616A..28Q} as long as the angular momentum transport within the BH progenitor star is efficient. The Tayler-Spruit (TS) dynamo proposed by \cite{2002A&A...381..923S}, is one of the well-accepted, efficient mechanisms for angular momentum transport inside stars. The TS dynamo is believed to be produced by differential rotation in the radiative layers. The evidence that the TS dynamo is supported can be found as follows. First, the flat rotation profile of the Sun is well reproduced with the TS dynamo \citep{2005A&A...440L...9E}. Second, stellar models with the TS dynamo can explain the observations of the spin of white dwarfs and neutron stars \citep{2005ApJ...626..350H,2008A&A...481L..87S,2020ApJ...902L..12Z}. It was then found by \cite{2012A&A...544L...4E} that models with only TS dynamo cannot produce the slow core rotation rates of red giants. A better agreement with asteroseismic measurements for lower core rotation rates of subgiants, can be predicted with a revised TS dynamo \citep{2019MNRAS.485.3661F} when compared with the original TS dynamo. \cite{2019A&A...631L...6E}, however, found that the revised prescription still encounters difficulties in producing the observed core rotation rates for red giant stars.

Recent measurements of LIGO and Virgo have shown that the inspiral effective spins $\chi_{\rm eff}$ are typically small. This is well explained by the isolated binary evolution channel \citep{2018A&A...616A..28Q,2020A&A...635A..97B,2020A&A...636A.104B} in which the TS dynamo is assumed. However, this assumption is not supported when applied to another BH binary system, i.e., BH high mass X-ray binary (HMXB). For the three HMXBs (Cygnus X-1, M33 X-7 and LMC X-1), the BH has been continuously found rotating extremely fast \footnote{We note that this is not fully accepted for the accuracy of BH spin estimates in X-ray binaries in the community, but the discussion of the uncertainties is beyond the scope of this work.}. It is believed that the BH spin has to be natal when considering the limited lifetime of the BH companion and assuming the Eddington-limited accretion rate onto the BH after its birth.

\cite{2010Natur.468...77V} proposed a so-called Case-A mass transfer channel \citep{1967ZA.....65..251K} that is applicable to the formation of M33 X-7. In this channel, the two stars evolve initially in a close binary system, and the BH progenitor star, while still in its Main Sequence, initiates mass transfer onto its companion. \cite{2019ApJ...870L..18Q} recently investigated this channel and found that, in order to explain the fast BH spin of the three BH HMXBs, a less efficient angular momentum transport mechanism than the TS dynamo is required. This indicates that a very strong angular momentum transport does not apply in every situation.

GW190412 reported recently by LIGO/Virgo \citep{2020PhRvD.102d3015A} is the first unequivocally unequal masses, mass ratio $q$ = $0.25^{+0.06}_{-0.04}$ (using the EOBNR PHM approximation). It was reported that this event is consistent with a moderately spinning primary orbiting around a secondary whose spin is unconstrained by using the uniform prior for individual spin magnitudes and isotropic directions, i.e., uninformative prior (LVC \textit{DEFAULT} prior). Instead, \cite{2020ApJ...895L..28M} motivated by the well-accepted astrophysical spin prior (namely, the primary spin is zero under the assumption of the efficient angular momentum transport), found that the secondary had a dimensionless spin component at least 0.64 (95\% confidence) along the orbital angular momentum. However, based on various spin prior assumptions, \cite{2020ApJ...899L..17Z} found that the non-spinning primary is disfavored by the data. Furthermore, a recent population synthesis study in \cite{2020ApJ...901L..39O} shows the binary evolution with the typical assumptions adopted before can reproduce the GW190412-like system.

This work is motivated by the recently reported candidate event in GWTC-2.1, GW190403$_{-}$051519 with very unequal masses and extremely high effective spin. If this event is astrophysical and formed via the classical isolated binary evolution channel, the conventional efficient angular momentum transport mechanism, (i.e., TS dynamo) will be significantly challenged. The layout of this work is organized as follows. In Section 2, we present the inspiral effective spins predicted by the isolated field binary evolution. The current BH spin measurements from LIGO/Virgo are discussed in Section 3. We then describe in Section 4 the comparisons between model predictions and observations. The discussions and conclusions are given in Section 5 and 6, respectively.

\section{The effective inspiral spins predicted by the isolated field binary evolution}
In the classical formation channel of binary BHs (BBHs), the angular momentum of their progenitors, dominated by different physical processes, results in the angular momentum of the BH and thus the BH spins. Here we briefly describe how the two BH spins are respectively determined (also see Section 2 in \cite{2020ApJ...895L..28M} for a longer discussion) and more detailed studies can be found in \cite{2018A&A...616A..28Q}.

\subsection{The primary spin}
First, the progenitor of the primary BH (the more massive star) evolves at a wide orbit in which the tides are too weak to change the spins of both components. Second, the stellar winds strip its outer layers and thus slow down the progenitor. Furthermore, the mass transfer, if it occurs after the Main Sequence phase, can also remove material from the primary to the secondary via the point $L_1$. The masses lost from the primary carry the corresponding specific angular momentum of its surface, and thus slows it down. In addition, a small fraction of mass accreted onto the primary during the CE phase is too small to spin up the BH \citep{2015ApJ...798L..19M}. Therefore, in this evolutionary scenario, combining the three potential processes indicates that the angular momentum of the primary's core, which will form the first-born BH, is mainly determined by the efficiency of the coupling between the stellar core and its envelope. 

A systematic investigations of single massive stars to directly collapse to form BHs, carried out recently by \cite{2018A&A...616A..28Q} with different initial conditions, metallicities, rotation rates and masses, show that the spins of the resultant first-born BHs are negligible. This statement, however, is made under the assumption of efficient angular momentum transport within the star. The TS dynamo can efficiently transport the angular momentum from the core to the envelope during the expansion and thus form a non-spinning BH at its birth. More recently, it was found that massive stars produce slowly spinning BHs with the TS dynamo \citep[a $\sim$ 0.1; see][]{2020A&A...636A.104B} and almost non-spinning BHs with more efficient TS dynamo \citep[a $\sim$ 0.01; see][]{2019MNRAS.485.3661F}, respectively.

\subsection{The secondary spin}
Following the formation of the first-born BH, its companion expands significantly after the Main Sequence phase. The binary system, due to the unequal mass ratio and expansion of the BH companion, then undergoes dynamically unstable mass transfer phase (also called CE phase). During this CE phase, The orbit shrinks dramatically by converting its orbital energy into heat. Eventually the post-CE system consists of a close binary, a helium star orbiting around a BH at a orbital period of a few days. Alternatively, the system may undergo stable mass transfer by Roche lobe overflow \citep{2017MNRAS.471.4256V,2017MNRAS.465.2092P,2019ApJ...882..121S,2019MNRAS.490.3740N,2021ApJ...920...81S} and allow to form a highly spinning primary BH.

\cite{2018A&A...616A..28Q} for the first time performed a systematic study of the evolution of the angular momentum of solid-body helium star with various initial conditions. They found that the angular momentum of the helium star is determined by the interplay of the stellar winds and the tides. If the tides are dominant over the winds mass-loss, a helium star can be synchronised with its companion in a close orbit, which results in a fast-spinning second-born BH orbiting around the non-spinning primary BH. Further study in \cite{2020A&A...635A..97B} shows that the TS dynamo within the helium star has a negligible impact on the spin of the second-born BH. A more recent study in \cite{2021ApJ...921L...2O} found that the isolated binary evolution can form close Wolf-Rayet star (helium star) + Wolf-Rayet star with an equal mass after the CE phase and thus tend to form fast spinning BHs. But this would not explain BBHs with very unequal masses.

\subsection{The effective spin $\chi_{\rm eff}$}\label{chi_eff}
\begin{figure}[htbp!]
     \centering
     \includegraphics[width=\columnwidth]{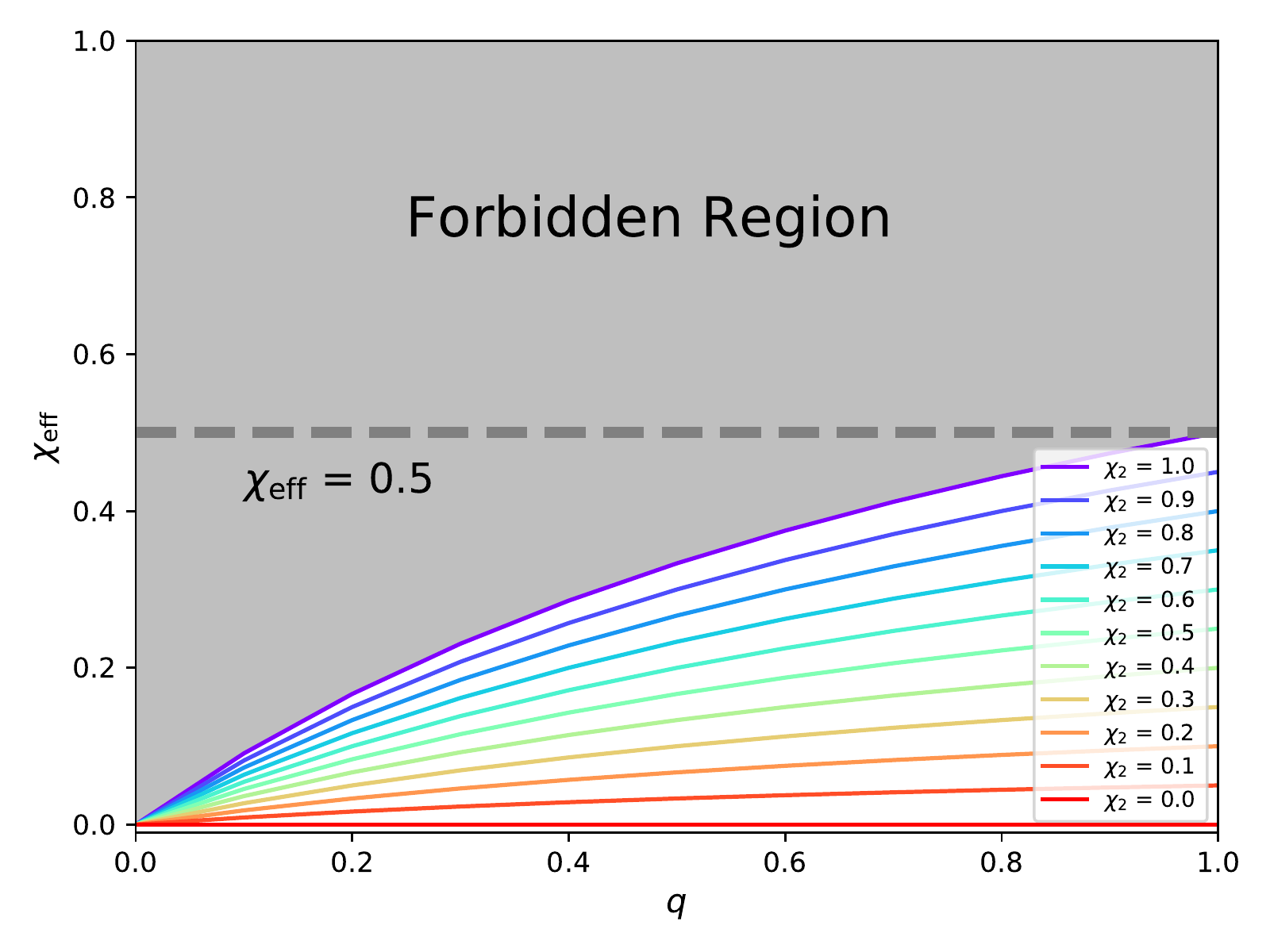}
     \caption{$\chi_{\rm eff}$ as a function of the mass ratio $q$ for a given primary spin $\chi_1$ = 0. Different colors refer to various secondary dimensionless spins $\chi_2$. The horizontal dashed line marks $\chi_{\rm eff}$ = 0.5. The grey area is the ``Forbidden Region'' in which both the $\chi_{\rm eff}$ and $q$ cannot be reached.}
     \label{a2}
\end{figure}

In the scenario of the classical field binary evolution formation channel for BBHs, it is believed in general that the two BHs are aligned to each other and to the total orbital angular momentum. Therefore, the effective inspiral spins are exclusively determined by the magnitude of the two BH spins. Assuming the highly efficient angular momentum transport \citep[e.g.,][]{2019MNRAS.485.3661F} within the primary star (more massive star) and no mass ratio reversal in the binary evolution, then the initial primary (more massive star on zero-age main sequence)
forms the first-born BH with $\vec{\chi_1} \sim \vec{0}$, and thus the effective spin is given as 
\begin{equation}\label{xeff}
    \chi_{\rm eff} = \frac{M_2 \vec{\chi_2}}{M_1+M_2}\cdot \hat{L}_N = \frac{q}{1+q} \vec{\chi_2} \cdot \hat{L}_N,
\end{equation}
where $q = M_2 / M_1$ is the mass ratio. The colored lines in Fig.~\ref{a2} show the effective spin $\chi_{\rm eff}$ as a function of the mass ratio $q$ for various secondary spins $\chi_2$, from non-spinning to spinning maximally. Under the assumption of the astrophysically-motivated primary spin, we can see that the $\chi_{\rm eff}$ increases with the mass ratio $q$, and that the $\chi_{\rm eff}$ has an upper limit of 0.5. The grey region marks the parameter space in which the effective $\chi_{\rm eff}$ and mass ratio $q$ cannot be reached (namely, the ``Forbidden Region''. Note that this is based on the assumption of non-spinning first-born BHs formed from the more massive stars in the classical isolated binary evolution channel, in which a close binary system after the common envelope phase is composed of a BH and a helium star.

\section{Current spin measurements from LIGO/Virgo}
\begin{figure*}
     \centering
     \includegraphics[width=0.99\textwidth]{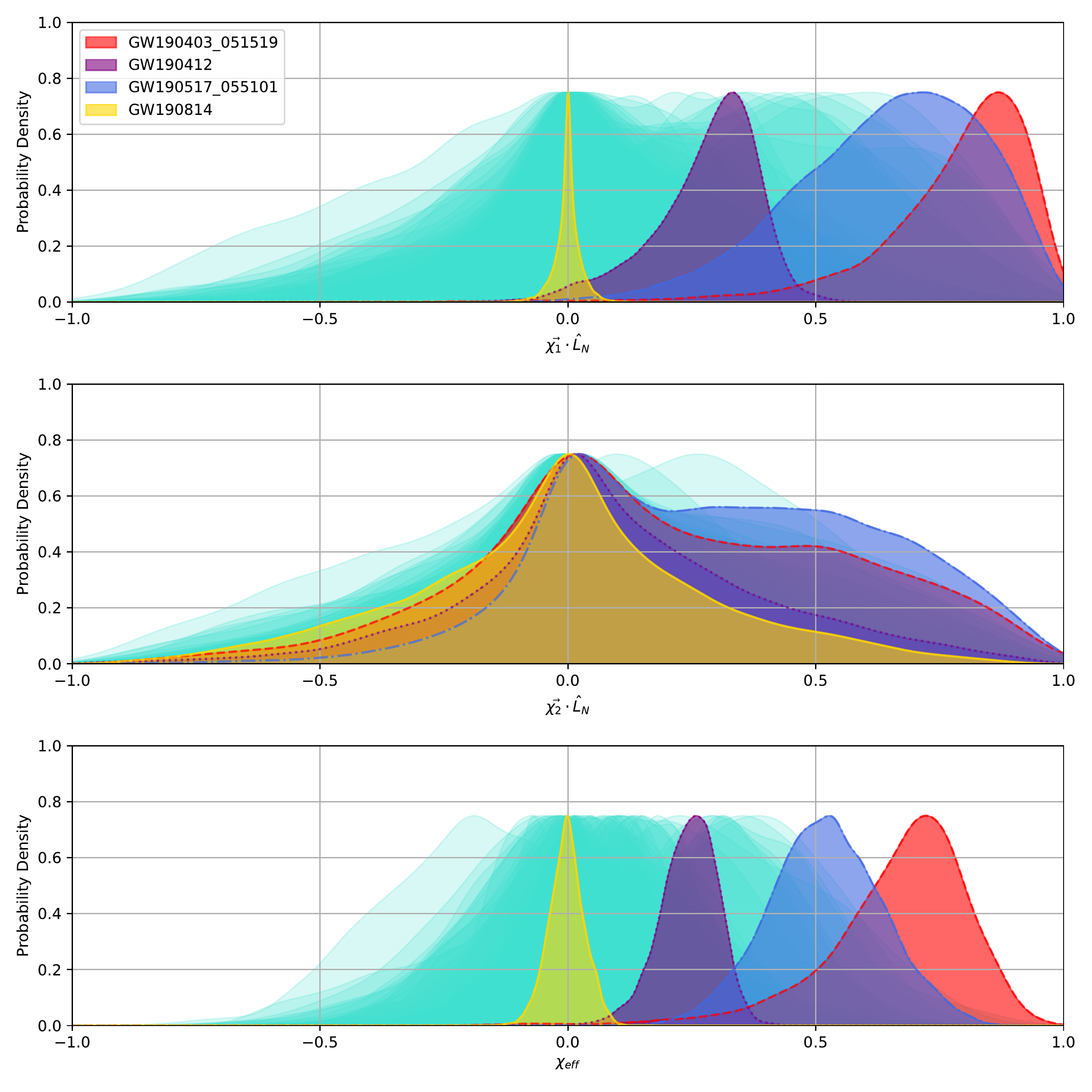}
     \caption{Posterior probability distributions of $\vec{\chi_1} \cdot \hat{L}_N$, $\vec{\chi_2} \cdot \hat{L}_N$ and $\chi_{\rm eff}$ for 54 BBHs from LIGO/Virgo . The turquoise represents 50 GW events, and 4 cases are highlighted with different colors (red for GW190403$_{-}$0501519, purple for GW190412, blue for GW190517$_{-}$055101, and golden for GW190814). }
     \label{x123}
\end{figure*}

Figure~\ref{x123} shows the probability density of the posterior distribution for the spin measurements ($\vec{\chi_1} \cdot \hat{L}_N$, $\vec{\chi_2} \cdot \hat{L}_N$, and $\chi_{\rm eff}$) based on the 54 LIGO/Virgo BBH candidates \footnote{We assume that GW190917$_{-}$114630 in GTWC-2.1 is a black hole-neutron star event.}. The posterior samples are obtained from the Gravitational Wave Open Science Center\footnote{\url{https://www.gw-openscience.org} and \url{https://zenodo.org/record/5117703}.}, with 4 individual cases (GW190412, GW190814 \footnote{We assume that GW190814 is a binary BH event.}, GW190517$_{-}$055101 and GW190403$_{-}$051519) highlighted in different colors. The highlighted events have extreme mass ratios, and/or large primary spins. We adopt the ``Overall$_{-}$posterior'', ``PublicationSamples'' and ``IMRPhenomXPHM$_{-}$comoving'' samples for the released data in GWTC-1, GWTC-2 and GWTC-2.1, respectively. Note that the parameter estimation processes to derive posterior samples are performed by the LVC  using the \textit{DEFAULT} prior (uniform in component masses, component spin magnitudes and isotropic spin directions).

The upper panel in Fig.~\ref{x123} shows the distribution for the projection of primary spins ($\vec{\chi_1} \cdot \hat{L}_N$) on the direction of the orbital angular momentum. For a substantial portion of events, their posterior distributions are symmetrically peaked at zero with similar shapes. This is because the magnitudes of their primary spin are loosely constrained, and thus the corresponding distributions are mainly determined by the prior. We note that a non-negligible fraction of events show posterior peaks extending over $\sim 0.5$, indicating that the BBHs, at least for some events, show primary BH spinning fast.

Four atypical events are highlighted in different colors. GW190412 is the first GW event with unequivocally unequal masses. This system has mass ratio $q = 0.28_{-0.07}^{+0.12}$ \citep{2020PhRvD.102d3015A}. It was reported that the primary rotated with a dimensionless spin between 0.22 and 0.6 (90\% probability). GW190814 is another interesting event \citep{2020ApJ...896L..44A} that has the smallest mass ratio. But the primary BH was reported to have negligible spin. GW190517$_{-}$055101 has the largest $\chi_{\rm eff}$ reported in GWTC-2. More recently, GW190403$_{-}$051519 in GWTC-2.1 shows support for high spin and unequal mass ratio ($q = 0.25_{-0.11}^{+0.54}$). The primary dimensionless spin is reported to be $\chi_1 = 0.92^{+0.07}_{-0.22}$. This marks the most extremely spin observed using GWs. Note that the primary spin magnitudes of GW190517$_{-}$055101 and GW190403$_{-}$0501519 are found to be $> 0.1$ over 99\% credibility.

In the middle panel of Fig.~\ref{x123}, the distribution of the secondary spins shows a dominant Gaussian component with a median around zero. The $\chi_{\rm eff}$ shown in the bottom panel has a similar shape of the primary spins, but with narrower extensions.

\section{Model predictions and observations}
It has been pointed out in \cite{2013PhRvD..87b4035B} that in the post-Newtonian theory, there is a well known degeneracy between mass ratio and total spin, which impairs the accurate extraction of the physical parameters. Therefore, the combination of $\chi_{\rm eff}$ and mass ratio $q$ can be used as a probe for comparing the observation with the theoretical prediction. 

\begin{figure*}
     \centering
     \includegraphics[width=0.9\textwidth]{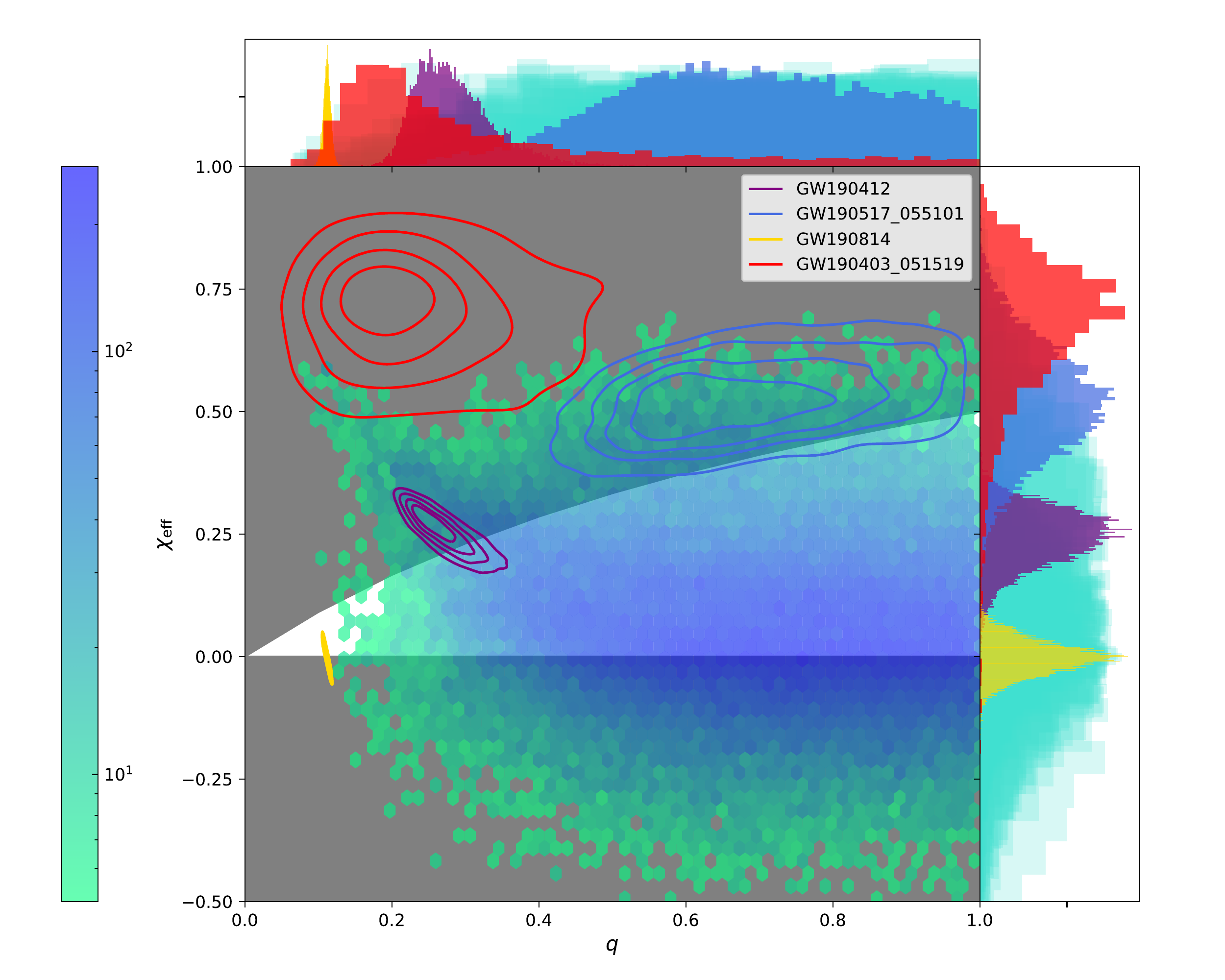}
     \caption{Observation for the $\chi_{\rm eff}$ versus mass ratio $q$ of 54 BBH events, with the theoretical ``Forbidden Region'' in grey as the background. The contours represent the posterior distributions for GW190412 (purple), GW190517$_{-}$055101 (blue), GW190814 (yellow) and GW190403$_{-}$0501519 (red), with their outer edges denoting the 68\% credible regions. The hexagonal binning plot shows the stacked posteriors of the 50 events (with 2000 randomly selected samples for each event), and its corresponding colorbar is shown on the left. One-dimensional histograms of 50 events are shown in turquoise, with the other 4 events in different colors.}
     \label{obs}
\end{figure*}

Figure~\ref{obs} shows for observed 54 BBHs the posterior distributions in the plane $\chi_{\rm eff} -q$ inferred by the the LVC  using the \textit{DEFAULT} prior, with the ``Forbidden Region'' shown in grey as the background. The contours of the posterior distributions for 4 events (GW190403$_{-}$0501519, GW190412, GW190517$_{-}$055101, and GW190814) can quantitatively illustrate how likely they lie inside the ``Forbidden Region''. For other 50 BBHs, their posterior samples are represented by the color of the filled hexagons. In addition to Fig.~\ref{obs}, in order to quantify how likely each event falls in the ``Forbidden Region'', we count the fraction of posterior samples $P_{\rm in}$ in this region, and the corresponding results are shown in Table~\ref{tb:pout}. We then highlight our main findings as follows.

Let us first discuss the sample of 50 events where the four events put in evidence in the figure are considered apart (see below). First of all, we can see from the Fig.~\ref{obs} that for the majority of the BHHs, the $\chi_{\rm eff}$ is clustered at $\sim 0$ and the mass ratio $q$ is in the range of $0.3<q<1$. Detailed population analysis in \citet{2021ApJ...913L...7A} shows that the $\chi_{\rm eff}$ distribution can be described as a Gaussian distribution peaked at a positive and small value ($\sim 0.06$), with a standard deviation of $\sim 0.12$. They also claimed that there is some evidence that a nonzero fraction of BBH systems show a negative effective inspiral spin parameter. However, \citet{2021PhRvD.104h3010R} argued that the result of negative $\chi_{\rm eff}$ is disfavored by the data.

We note that the probability density of merging BBH events coming from extreme mass ratio ($q<0.2$) is very low, although the distribution of mass ratio is found to be broad \citep{2021ApJ...913L...7A}. Moreover, we find that all of these 50 events have $P_{\rm in}$ less than 0.9, and among them 16 events have $P_{\rm in} > 0.5$. GW190828$_{-}$063405 and GW190514$_{-}$065416 have the smallest and largest $P_{\rm in}$ (0.03 and 0.85), respectively.

Next, let us now discuss the four cases standing out by either $\chi_{\rm eff}$ or mass ratio $q$ value. In the current BBH populations reported by the LVC, the number of events with very unequal masses is very limited. For the two extreme mass ratio events, GW190814 and GW190412, we find that they have a significant parts of their posteriors in the ``Forbidden Region''. Additionally, GW190517$_{-}$055101 was reported with the largest $\chi_{\rm eff}$ in GWTC-2, but is consistent with a typical mass ratio \citep{2021PhRvX..11b1053A}. These three events are all more likely to lie in the ``Forbidden Region'', but none of them has a $P_{\rm in} > 0.9$. Among the recently reported events in GWTC-2.1, GW190403$_{-}$051519 is the only event with its $68\%$ credible region inside the ``Forbidden Region'' and it has the largest $P_{\rm in}$ (0.96) among all events, making it the most outstanding event ever observed.

\begin{figure*}
     \centering
     \includegraphics[width=0.9\textwidth]{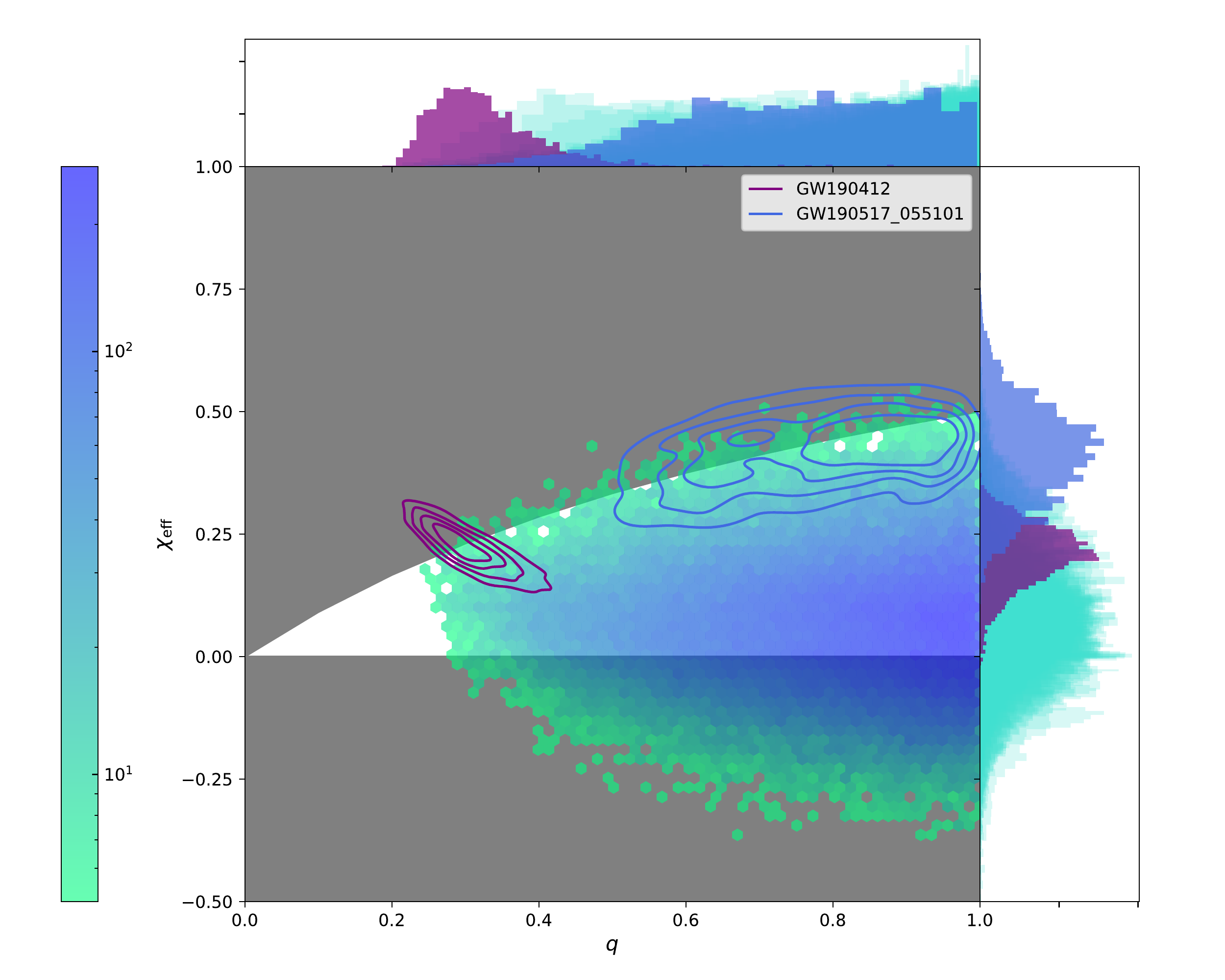}
     \caption{The same as Fig.~\ref{obs}, but for the posterior samples re-weighted by the population informed prior. The two ``outliers", GW190814 and GW190403$_{-}$0501519, are not included.}
     \label{obs_pop}
\end{figure*}

Recent studies have shown that the population informed priors have non-negligible impacts on the inferred mass ratio and $\chi_{\rm eff}$ for some events \citep{2020ApJ...904L..26F,2020ApJ...895..128M,2021arXiv210600521C}. To derive the re-weighted posterior distributions, for each event, following \citet{2020ApJ...895..128M} we select 5000 random parameter estimation samples $\{\chi_{\rm eff},m_1,m_2,z \}_i$ (corresponding to the effective spins, component masses and redshift) subjected to the weights:
 \begin{equation}\label{weight} 
\mathit{w}_i = \frac{p_{\rm pop}(\chi_{{\rm eff},i})p_{
\rm pop}(m_{1,i},m_{2,i},z_i)}{p_{\rm pe}(\chi_{{\rm eff},i})p_{
\rm pe}(m_{1,i},m_{2,i},z_i)}, 
\end{equation}
where $p_{\rm pop}$ and $p_{\rm pe}$ represent the probability density of population informed prior and the \textit{DEFAULT} prior used for each event's parameter estimation, respectively. For $p_{\rm pop}(\chi_{{\rm eff}})$, we utilize the posterior predictive distribution of the GAUSSIAN spin model, while for $p_{\rm pop}(m_1,m_2,z_i)$, we adopt the result from the POWER LAW + PEAK model. Similar to Fig.~\ref{obs}, we present the re-weighted posterior distributions using the population informed prior in Fig.~\ref{obs_pop}. Nevertheless, two events (GW190814 and GW190403$_{-}$051519), are excluded from the re-weighting analysis. As found by \citet{2021ApJ...913L...7A}, the population of GWTC-2 BBHs prefers events with low masses, nearly equal mass ratios and small $\chi_{\rm eff}$. They pointed out that GW190814 is an outlier in both secondary mass and mass ratio. As for GW190403$_{-}$051519, there are $48\%$ of its primary mass samples larger than the constrained median of the maximum mass ($m_{\rm max}$) for POWER LAW + PEAK model in \citet{2021ApJ...913L...7A}, $80\%$ of its mass ratio samples smaller than 0.5, and $88\%$ of its $\chi_{\rm eff}$ samples larger than 0.5. Thus similar to GW190814, the population informed prior is not likely to be suitable for GW190403$_{-}$051519 either. We also present the fraction ($P_{\rm in,pop}$) of the re-weighted posterior samples inside the ``Forbidden Region'' in Table~\ref{tb:pout} for the rest 52 events. When compared with the result derived from the \textit{DEFAULT} prior, the number of events with $P_{\rm in} > 0.5$ for the 50 BBHs is significantly decreased down to 5.

\begin{table*}[t]
	\centering
	\caption{The Fraction of Posterior Samples Inside the ``Forbidden Region'' for Each Source.}
	\begin{ruledtabular}
		\label{tb:pout}
		\begin{tabular}{ccccccccc}
Event & $P_{\rm in}$ & $P_{\rm in,pop}$ & Event & $P_{\rm in}$ & $P_{\rm in,pop}$ & Event  & $P_{\rm in}$ & $P_{\rm in,pop}$\\ \hline
GW150914 & 0.57 & 0.45 & GW151012 & 0.46 & 0.35 & GW151226 & 0.26 & 0.087 \\
GW170104 & 0.61 & 0.44 & GW170608 & 0.33 & 0.33 & GW170729 & 0.4 & 0.053 \\
GW170809 & 0.28 & 0.21 & GW170814 & 0.27 & 0.17 & GW170818 & 0.73 & 0.54 \\
GW170823 & 0.32 & 0.23 & GW190408$_{-}$181802 & 0.66 & 0.49 & GW190413$_{-}$052954 & 0.54 & 0.35 \\
GW190413$_{-}$134308 & 0.56 & 0.37 & GW190421$_{-}$213856 & 0.67 & 0.48 & GW190424$_{-}$180648 & 0.18 & 0.13 \\
GW190503$_{-}$185404 & 0.61 & 0.45 & GW190512$_{-}$180714 & 0.36 & 0.34 & GW190513$_{-}$205428 & 0.24 & 0.15 \\
GW190514$_{-}$065416 & 0.85 & 0.62 & GW190519$_{-}$153544 & 0.3 & 0.052 & GW190521 & 0.47 & 0.28 \\
GW190521$_{-}$074359 & 0.14 & 0.087 & GW190527$_{-}$092055 & 0.31 & 0.18 & GW190602$_{-}$175927 & 0.32 & 0.2 \\
GW190620$_{-}$030421 & 0.36 & 0.044 & GW190630$_{-}$185205 & 0.11 & 0.075 & GW190701$_{-}$203306 & 0.69 & 0.5 \\
GW190706$_{-}$222641 & 0.34 & 0.088 & GW190707$_{-}$093326 & 0.8 & 0.92 & GW190708$_{-}$232457 & 0.3 & 0.3 \\
GW190719$_{-}$215514 & 0.46 & 0.098 & GW190909$_{-}$114149 & 0.68 & 0.45 & GW190720$_{-}$000836 & 0.14 & 0.0082 \\
GW190727$_{-}$060333 & 0.26 & 0.16 & GW190728$_{-}$064510 & 0.11 & 0.0024 & GW190731$_{-}$140936 & 0.35 & 0.22 \\
GW190803$_{-}$022701 & 0.6 & 0.41 & GW190828$_{-}$063405 & 0.031 & 0.02 & GW190828$_{-}$065509 & 0.26 & 0.28 \\
GW190910$_{-}$112807 & 0.41 & 0.32 & GW190915$_{-}$235702 & 0.43 & 0.26 & GW190924$_{-}$021846 & 0.44 & 0.81 \\
GW190929$_{-}$012149 & 0.65 & 0.31 & GW190930$_{-}$133541 & 0.22 & 0.091 & GW190426$_{-}$190642 & 0.48 & 0.18 \\
GW190725$_{-}$174728 & 0.65 & 0.78 & GW190805$_{-}$211137 & 0.47 & 0.089 & GW190916$_{-}$200658 & 0.35 & 0.15 \\
GW190925$_{-}$232845 & 0.18 & 0.1 & GW190926$_{-}$050336 & 0.63 & 0.41 & GW190412 & 0.63 & 0.38 \\
GW190517$_{-}$055101 & 0.87 & 0.43 & GW190814 & 0.53 & -- & GW190403$_{-}$051519 & 0.96 & -- \\
		\end{tabular}
		\tablenotetext{}{$P_{\rm in}$ are calculated using the original PE posterior samples, while $P_{\rm in,pop}$ are derived using the samples re-weighted by the population informed prior.}
	\end{ruledtabular}
\end{table*}

\section{Discussions}
Recent predictions \citep{2019MNRAS.490.3740N,2020A&A...636A.104B,2020A&A...635A..97B,2021A&A...647A.153B,2021MNRAS.505..339M,2021arXiv210505783G}, had been in a good agreement with the measurements of BBHs by the end of the O3a data release. All these detailed binary evolution calculations are based on the efficient angular momentum transport (i.e., TS dynamo), which is still one of main uncertainties in stellar evolution. Stellar models with the TS dynamo mechanism can well match the rotation rates for the Sun, white dwarfs, as well as neutron stars. However, a new detection of GW190403$_{-}$0501519 reported in GTWC-2.1 with a very unequal mass ratio $q$ and the highest magnitude of $\chi_{\rm eff}$, is in favor of the primary BH spinning extremely fast. This result implies that the conventional angular momentum transport mechanism in the classical isolated binary evolution channel is facing an unprecedented challenge. The physics of the angular momentum transport inside the star is still unclear. In cases the angular momentum transport is mainly dominated by meridional currents, a moderate coupling allows the stellar core to keep more angular momentum and thus produces fast-spinning neutron stars and BHs \citep{2012A&A...542A..29G}. \cite{2019ApJ...870L..18Q} recently found that weak coupling between the core and envelope inside the star after its Main Sequence phase is required to explain the BH high spins for Cygnus X-1, M33 X-7 and LMC X-1. 

Furthermore, mass loss for massive stars is highly uncertain. The new observations of Cygnus X-1 \citep{2021Sci...371.1046M} indicates the currently used winds prescription for massive stars are overestimated. \cite{2020MNRAS.499..873S} found the wind mass-loss rate from recent theoretical modeling for stripped stars is weaker than previous predictions. Therefore, single fast-rotating massive stars with winds reduced could retain more angular momentum and thus favor chemically homogeneous evolution. Note that massive stars at high initial rotation rates or close binaries evolve chemically homogeneously, but still form non-spinning BHs when considering efficient angular momentum transport \cite[see Fig. 3 in][]{2019ApJ...870L..18Q}. 

Binary BH formation through chemically homogeneous evolution \citep{2016MNRAS.458.2634M,2016MNRAS.460.3545D,2019MNRAS.490.3740N} produces two fast-spinning BHs with nearly equal masses. Therefore, GW190403$_{-}$0501519 is unlikely to have formed through this channel. The reported primary BH mass ($m_1 = 88.0_{-32.9}^{+28.2} M_{\odot}$) of this event falls in the mas gap predicted by pair-instability theory \citep{1964ApJS....9..201F,2003ApJ...591..288H}. However, the boundary of the mass gap is still theoretically uncertain, and is sensitive to detailed stellar evolution, including uncertainties on nuclear reaction rates \citep{2020ApJ...902L..36F}, metallicity \citep{2020ApJ...900...98G,2021MNRAS.502L..40F,2021MNRAS.504..146V}, rotation \citep{2020A&A...640L..18M}, etc. In addition, the BH with its mass in this gap could also be formed from Population III binary star evolution \citep{2021MNRAS.501L..49K}, or from hyper-Eddington accretion after black hole birth \citep{2021ApJ...912L..31W}, or from an earlier study of binary evolution of normal stars through the CE phase \citep{2020ApJ...905L..15B}. We note that the inferred two BH masses of GW190521 could be dependent on the \textit{DEFAULT} prior on $m_1$ and $m_2$ adopted in \cite{2021arXiv210801045T}, assuming BH detector-frame masses are uniformly distributed. By using a population-informed prior, \cite{2020ApJ...904L..26F} argued that GW190521 might be a straddling binary BHs, with two masses outside of this mass gap. Thus we suggest that a similar investigation for GW190403$_{-}$0501519 should be carried out. Alternatively, GW190403$_{-}$0501519 could contain two second-generation BHs, similar to GW190521 as in \cite{2021ApJ...915L..35K}.

Recent population synthesis study in \cite{2020ApJ...901L..39O} shows that 9.2\% of merging BBH systems in the local Universe (z $\sim$ 0) are expected to have mass ratio $q$ less than 0.41, which is consistent with the 90\% upper limit on the mass ratio of GW190412. Furthermore, we also note in Fig. 5 that the effective spins $\chi_{\rm eff}$ of BBH mergers can reach 1.0, while for mass ratio $q <$ 0.41, the effective spins $\chi_{\rm eff}$ are limited to $\sim$ 0.5. In a more systematic investigation for the BBHs reported in GWTC-2, \cite{2021ApJ...910..152Z} studied a mixture channel of isolated binary evolution and dynamical formation in globular clusters, and then concluded that any single channel does not contribute to more than $\simeq$ 70\% of the whole sample. However, \cite{2021arXiv210810885B} demonstrated that drawing conclusion about the quantitative comparison between formation channels for currently observed BBH populations cannot be robust due to physics uncertainties. 

More recently, \cite{2021ApJ...921L...2O} proposed two alternative binary evolution channels for producing BBHs with high effective spins. For the first scenario (see Fig. 1), the system has gone through the stable mass transfer instead of the CE phase, and the less component gains enough mass to first form the more massive BH with the spin of 0.68. In the second scenario (see Fig. 2), the equal-mass helium stars are formed after the CE phase and both BHs have the spin of 0.79. The two scenarios proposed by \cite{2021ApJ...921L...2O} could also be responsible for producing fast-spinning BBHs, for instance the case of GW190403$_{-}$051519. However, the two scenarios are different from the classical isolated binary evolution channel we study in this work.

\section{Conclusions}
In this work, we present for the classical isolated binary evolution channel (i.e., CE channel) of BBH formation the theoretical ``Forbidden Region'', in which the $\chi_{\rm eff}$ and $q$ cannot be reached. We here highlight that this channel specifically refers to the channel in which a BH and a helium star is formed immediately following the CE phase. Therefore, this result, based on the CE channel after which a close binary system of a BH produced by the more massive star and a helium star is formed, is exclusively dependent on the astrophysically motivated assumption of the primary BH spin being negligible, which is determined by the efficient angular momentum transport within the star. \cite{2021arXiv210600521C} Recently found an anti-correlation between the $\chi_{\rm eff}$ and mass ratio $q$ at 98.7\% credibility, showing more unequal-mass BBHs in favor of statistically larger $\chi_{\rm eff}$. With more BBHs reported recently in GWTC-3 \citep{2021arXiv211103606T}, this anti-correlation is confirmed at a high credibility \citep[see Fig. 21 in ][]{2021arXiv211103634T}. This finding, however, is generally inconsistent with the predictions of the classical isolated binary evolution channel under the assumption of the efficient angular momentum transport inside massive stars.

We then investigate all the BH spin measurements of BBHs that have been reported by LIGO/Virgo to date. GW190403$_{-}$0501519 has stood out from all other measurements by its very low mass ratio and its high inspiral effective spin. The detection of GW190403$_{-}$0101519, if formed through the classical isolated binary evolution channel, will definitely pose an unprecedented challenge for the conventional angular momentum transport mechanism of massive stars. We expect that more events like GW190403$_{-}$0501519 will be reported in the upcoming O3b, with which stronger constraints on the angular momentum transport efficiency within the star can be made.

\begin{acknowledgements}
We thank Christopher Berry for helpful comments on the manuscript. YQ acknowledges the support from the Doctoral research start-up funding of Anhui Normal University and the funding from Key Laboratory for Relativistic Astrophysics in Guangxi University. Dong-Hong Wu acknowledges the support from the Doctoral research start-up funding of Anhui Normal University and the funding from Key Laboratory of Modern Astronomy and Astrophysics in Ministry of Education, Nanjing University. GM has received funding from the European Research Council (ERC) under the European Union's Horizon 2020 research and innovation programme (grant agreement No 833925, project STAREX). Hangfeng Song is supported by the National Natural Science Foundation of China(GrantNos.11863003, 12173010).

All figures are made with the free Python module Matplotlib \citep{2007CSE.....9...90H}. We would also like to thank all of the essential workers who put their health at risk during the COVID-19 pandemic, without whom we would not have been able to complete this work.
\end{acknowledgements}


\clearpage

\begin{thebibliography}{}

\bibitem[Olejak \& Belczynski(2021)]{2021ApJ...921L...2O} Olejak, A. \& Belczynski, K.\ 2021, \apjl, 921, L2. doi:10.3847/2041-8213/ac2f48
\bibitem[The LIGO Scientific Collaboration et al.(2021b)]{2021arXiv211103634T} The LIGO Scientific Collaboration, The Virgo Collaboration, \& The KAGRA Scientific Collaboration\ 2021, arXiv:2111.03634
\bibitem[The LIGO Scientific Collaboration et al.(2021a)]{2021arXiv211103606T} The LIGO Scientific Collaboration, the Virgo Collaboration, the KAGRA Collaboration, et al.\ 2021, arXiv:2111.03606
\bibitem[Shao \& Li(2021)]{2021ApJ...920...81S} Shao, Y. \& Li, X.-D.\ 2021, \apj, 920, 81. doi:10.3847/1538-4357/ac173e
\bibitem[Roulet et al.(2021)]{2021PhRvD.104h3010R} Roulet, J., Chia, H.~S., Olsen, S., et al.\ 2021, \prd, 104, 083010. doi:10.1103/PhysRevD.104.083010
\bibitem[Belczynski et al.(2021)]{2021arXiv210810885B} Belczynski, K., Romagnolo, A., Olejak, A., et al.\ 2021, arXiv:2108.10885
\bibitem[The LIGO Scientific Collaboration et al.(2021)]{2021arXiv210801045T} The LIGO Scientific Collaboration, the Virgo Collaboration, Abbott, R., et al.\ 2021, arXiv:2108.01045
\bibitem[Mapelli et al.(2021)]{2021MNRAS.505..339M} Mapelli, M., Dall'Amico, M., Bouffanais, Y., et al.\ 2021, \mnras, 505, 339. doi:10.1093/mnras/stab1334
\bibitem[Abbott et al.(2021)]{2021ApJ...915L...5A} Abbott, R., Abbott, T.~D., Abraham, S., et al.\ 2021, \apjl, 915, L5. doi:10.3847/2041-8213/ac082e
\bibitem[Kimball et al.(2021)]{2021ApJ...915L..35K} Kimball, C., Talbot, C., Berry, C.~P.~L., et al.\ 2021, \apjl, 915, L35. doi:10.3847/2041-8213/ac0aef
\bibitem[Callister et al.(2021)]{2021arXiv210600521C} Callister, T.~A., Haster, C.-J., Ng, K.~K.~Y., et al.\ 2021, arXiv:2106.00521
\bibitem[Vink et al.(2021)]{2021MNRAS.504..146V} Vink, J.~S., Higgins, E.~R., Sander, A.~A.~C., et al.\ 2021, \mnras, 504, 146. doi:10.1093/mnras/stab842
\bibitem[Abbott et al.(2021)]{2021ApJ...913L...7A} Abbott, R., Abbott, T.~D., Abraham, S., et al.\ 2021, \apjl, 913, L7. doi:10.3847/2041-8213/abe949
\bibitem[Woosley \& Heger(2021)]{2021ApJ...912L..31W} Woosley, S.~E. \& Heger, A.\ 2021, \apjl, 912, L31. doi:10.3847/2041-8213/abf2c4
\bibitem[Ghodla et al.(2021)]{2021arXiv210505783G} Ghodla, S., van Zeist, W.~G.~J., Eldridge, J.~J., et al.\ 2021, arXiv:2105.05783
\bibitem[Zevin et al.(2021)]{2021ApJ...910..152Z} Zevin, M., Bavera, S.~S., Berry, C.~P.~L., et al.\ 2021, \apj, 910, 152. doi:10.3847/1538-4357/abe40e
\bibitem[Abbott et al.(2021)]{2021PhRvX..11b1053A} Abbott, R., Abbott, T.~D., Abraham, S., et al.\ 2021, Physical Review X, 11, 021053. doi:10.1103/PhysRevX.11.021053
\bibitem[Farrell et al.(2021)]{2021MNRAS.502L..40F} Farrell, E., Groh, J.~H., Hirschi, R., et al.\ 2021, \mnras, 502, L40. doi:10.1093/mnrasl/slaa196
\bibitem[Miller-Jones et al.(2021)]{2021Sci...371.1046M} Miller-Jones, J.~C.~A., Bahramian, A., Orosz, J.~A., et al.\ 2021, Science, 371, 1046. doi:10.1126/science.abb3363
\bibitem[Bavera et al.(2021)]{2021A&A...647A.153B} Bavera, S.~S., Fragos, T., Zevin, M., et al.\ 2021, \aap, 647, A153. doi:10.1051/0004-6361/202039804
\bibitem[Kinugawa et al.(2021)]{2021MNRAS.501L..49K} Kinugawa, T., Nakamura, T., \& Nakano, H.\ 2021, \mnras, 501, L49. doi:10.1093/mnrasl/slaa191
\bibitem[Belczynski(2020)]{2020ApJ...905L..15B} Belczynski, K.\ 2020, \apjl, 905, L15. doi:10.3847/2041-8213/abcbf1
\bibitem[Fishbach \& Holz(2020)]{2020ApJ...904L..26F} Fishbach, M. \& Holz, D.~E.\ 2020, \apjl, 904, L26. doi:10.3847/2041-8213/abc827
\bibitem[du Buisson et al.(2020)]{2020MNRAS.499.5941D} du Buisson, L., Marchant, P., Podsiadlowski, P., et al.\ 2020, \mnras, 499, 5941. doi:10.1093/mnras/staa3225
\bibitem[Sander \& Vink(2020)]{2020MNRAS.499..873S} Sander, A.~A.~C. \& Vink, J.~S.\ 2020, \mnras, 499, 873. doi:10.1093/mnras/staa2712
\bibitem[Farmer et al.(2020)]{2020ApJ...902L..36F} Farmer, R., Renzo, M., de Mink, S.~E., et al.\ 2020, \apjl, 902, L36. doi:10.3847/2041-8213/abbadd
\bibitem[Olejak et al.(2020)]{2020ApJ...901L..39O} Olejak, A., Fishbach, M., Belczynski, K., et al.\ 2020, \apjl, 901, L39. doi:10.3847/2041-8213/abb5b5
\bibitem[Zhu \& Ashton(2020)]{2020ApJ...902L..12Z} Zhu, X.-J. \& Ashton, G.\ 2020, \apjl, 902, L12. doi:10.3847/2041-8213/abb6ea
\bibitem[Groh et al.(2020)]{2020ApJ...900...98G} Groh, J.~H., Farrell, E.~J., Meynet, G., et al.\ 2020, \apj, 900, 98. doi:10.3847/1538-4357/aba2c8
\bibitem[Abbott et al.(2020)]{2020PhRvD.102d3015A} Abbott, R., Abbott, T.~D., Abraham, S., et al.\ 2020, \prd, 102, 043015. doi:10.1103/PhysRevD.102.043015
\bibitem[Zevin et al.(2020)]{2020ApJ...899L..17Z} Zevin, M., Berry, C.~P.~L., Coughlin, S., et al.\ 2020, \apjl, 899, L17. doi:10.3847/2041-8213/aba8ef
\bibitem[Marchant \& Moriya(2020)]{2020A&A...640L..18M} Marchant, P. \& Moriya, T.~J.\ 2020, \aap, 640, L18. doi:10.1051/0004-6361/202038902
\bibitem[Miller et al.(2020)]{2020ApJ...895..128M} Miller, S., Callister, T.~A., \& Farr, W.~M.\ 2020, \apj, 895, 128. doi:10.3847/1538-4357/ab80c0
\bibitem[Mandel \& Fragos(2020)]{2020ApJ...895L..28M} Mandel, I. \& Fragos, T.\ 2020, \apjl, 895, L28. doi:10.3847/2041-8213/ab8e41
\bibitem[Abbott et al.(2020)]{2020ApJ...896L..44A} Abbott, R., Abbott, T.~D., Abraham, S., et al.\ 2020, \apjl, 896, L44. doi:10.3847/2041-8213/ab960f
\bibitem[Belczynski et al.(2020)]{2020A&A...636A.104B} Belczynski, K., Klencki, J., Fields, C.~E., et al.\ 2020, \aap, 636, A104. doi:10.1051/0004-6361/201936528
\bibitem[Bavera et al.(2020)]{2020A&A...635A..97B} Bavera, S.~S., Fragos, T., Qin, Y., et al.\ 2020, \aap, 635, A97. doi:10.1051/0004-6361/201936204
\bibitem[Neijssel et al.(2019)]{2019MNRAS.490.3740N} Neijssel, C.~J., Vigna-G{\'o}mez, A., Stevenson, S., et al.\ 2019, \mnras, 490, 3740. doi:10.1093/mnras/stz2840
\bibitem[Eggenberger et al.(2019)]{2019A&A...631L...6E} Eggenberger, P., den Hartogh, J.~W., Buldgen, G., et al.\ 2019, \aap, 631, L6. doi:10.1051/0004-6361/201936348
\bibitem[Stevenson et al.(2019)]{2019ApJ...882..121S} Stevenson, S., Sampson, M., Powell, J., et al.\ 2019, \apj, 882, 121. doi:10.3847/1538-4357/ab3981
\bibitem[Abbott et al.(2019)]{2019PhRvX...9c1040A} Abbott, B.~P., Abbott, R., Abbott, T.~D., et al.\ 2019, Physical Review X, 9, 031040. doi:10.1103/PhysRevX.9.031040
\bibitem[Fuller et al.(2019)]{2019MNRAS.485.3661F} Fuller, J., Piro, A.~L., \& Jermyn, A.~S.\ 2019, \mnras, 485, 3661. doi:10.1093/mnras/stz514
\bibitem[Qin et al.(2019)]{2019ApJ...870L..18Q} Qin, Y., Marchant, P., Fragos, T., et al.\ 2019, \apjl, 870, L18. doi:10.3847/2041-8213/aaf97b
\bibitem[Qin et al.(2018)]{2018A&A...616A..28Q} Qin, Y., Fragos, T., Meynet, G., et al.\ 2018, \aap, 616, A28. doi:10.1051/0004-6361/201832839
\bibitem[van den Heuvel et al.(2017)]{2017MNRAS.471.4256V} van den Heuvel, E.~P.~J., Portegies Zwart, S.~F., \& de Mink, S.~E.\ 2017, \mnras, 471, 4256. doi:10.1093/mnras/stx1430
\bibitem[Abbott et al.(2017)]{2017PhRvL.119p1101A} Abbott, B.~P., Abbott, R., Abbott, T.~D., et al.\ 2017, \prl, 119, 161101. doi:10.1103/PhysRevLett.119.161101
\bibitem[Pavlovskii et al.(2017)]{2017MNRAS.465.2092P} Pavlovskii, K., Ivanova, N., Belczynski, K., et al.\ 2017, \mnras, 465, 2092. doi:10.1093/mnras/stw2786
\bibitem[de Mink \& Mandel(2016)]{2016MNRAS.460.3545D} de Mink, S.~E. \& Mandel, I.\ 2016, \mnras, 460, 3545. doi:10.1093/mnras/stw1219
\bibitem[Belczynski et al.(2016)]{2016Natur.534..512B} Belczynski, K., Holz, D.~E., Bulik, T., et al.\ 2016, \nat, 534, 512. doi:10.1038/nature18322
\bibitem[Mandel \& de Mink(2016)]{2016MNRAS.458.2634M} Mandel, I. \& de Mink, S.~E.\ 2016, \mnras, 458, 2634. doi:10.1093/mnras/stw379
\bibitem[Marchant et al.(2016)]{2016A&A...588A..50M} Marchant, P., Langer, N., Podsiadlowski, P., et al.\ 2016, \aap, 588, A50. doi:10.1051/0004-6361/201628133
\bibitem[Abbott et al.(2016)]{2016ApJ...818L..22A} Abbott, B.~P., Abbott, R., Abbott, T.~D., et al.\ 2016, \apjl, 818, L22. doi:10.3847/2041-8205/818/2/L22
\bibitem[Abbott et al.(2016)]{2016PhRvL.116f1102A} Abbott, B.~P., Abbott, R., Abbott, T.~D., et al.\ 2016, \prl, 116, 061102. doi:10.1103/PhysRevLett.116.061102
\bibitem[Rodriguez et al.(2015)]{2015PhRvL.115e1101R} Rodriguez, C.~L., Morscher, M., Pattabiraman, B., et al.\ 2015, \prl, 115, 051101. doi:10.1103/PhysRevLett.115.051101
\bibitem[LIGO Scientific Collaboration et al.(2015)]{2015CQGra..32g4001L} LIGO Scientific Collaboration, Aasi, J., Abbott, B.~P., et al.\ 2015, Classical and Quantum Gravity, 32, 074001. doi:10.1088/0264-9381/32/7/074001
\bibitem[Acernese et al.(2015)]{2015CQGra..32b4001A} Acernese, F., Agathos, M., Agatsuma, K., et al.\ 2015, Classical and Quantum Gravity, 32, 024001. doi:10.1088/0264-9381/32/2/024001
\bibitem[MacLeod \& Ramirez-Ruiz(2015)]{2015ApJ...798L..19M} MacLeod, M. \& Ramirez-Ruiz, E.\ 2015, \apjl, 798, L19. doi:10.1088/2041-8205/798/1/L19
\bibitem[Cantiello et al.(2014)]{2014ApJ...788...93C} Cantiello, M., Mankovich, C., Bildsten, L., et al.\ 2014, \apj, 788, 93. doi:10.1088/0004-637X/788/1/93
\bibitem[Ivanova et al.(2013)]{2013A&ARv..21...59I} Ivanova, N., Justham, S., Chen, X., et al.\ 2013, \aapr, 21, 59. doi:10.1007/s00159-013-0059-2
\bibitem[Baird et al.(2013)]{2013PhRvD..87b4035B} Baird, E., Fairhurst, S., Hannam, M., et al.\ 2013, \prd, 87, 024035. doi:10.1103/PhysRevD.87.024035
\bibitem[Eggenberger et al.(2012)]{2012A&A...544L...4E} Eggenberger, P., Montalb{\'a}n, J., \& Miglio, A.\ 2012, \aap, 544, L4. doi:10.1051/0004-6361/201219729
\bibitem[Georgy et al.(2012)]{2012A&A...542A..29G} Georgy, C., Ekstr{\"o}m, S., Meynet, G., et al.\ 2012, \aap, 542, A29. doi:10.1051/0004-6361/201118340
\bibitem[Valsecchi et al.(2010)]{2010Natur.468...77V} Valsecchi, F., Glebbeek, E., Farr, W.~M., et al.\ 2010, \nat, 468, 77. doi:10.1038/nature09463
\bibitem[Suijs et al.(2008)]{2008A&A...481L..87S} Suijs, M.~P.~L., Langer, N., Poelarends, A.-J., et al.\ 2008, \aap, 481, L87. doi:10.1051/0004-6361:200809411
\bibitem[Hunter(2007)]{2007CSE.....9...90H} Hunter, J.~D.\ 2007, Computing in Science and Engineering, 9, 90 
\bibitem[Eggenberger et al.(2005)]{2005A&A...440L...9E} Eggenberger, P., Maeder, A., \& Meynet, G.\ 2005, \aap, 440, L9. doi:10.1051/0004-6361:200500156
\bibitem[Heger et al.(2005)]{2005ApJ...626..350H} Heger, A., Woosley, S.~E., \& Spruit, H.~C.\ 2005, \apj, 626, 350. doi:10.1086/429868
\bibitem[Heger et al.(2003)]{2003ApJ...591..288H} Heger, A., Fryer, C.~L., Woosley, S.~E., et al.\ 2003, \apj, 591, 288. doi:10.1086/375341
\bibitem[Spruit(2002)]{2002A&A...381..923S} Spruit, H.~C.\ 2002, \aap, 381, 923. doi:10.1051/0004-6361:20011465
\bibitem[Tutukov \& Yungelson(1993)]{1993MNRAS.260..675T} Tutukov, A.~V. \& Yungelson, L.~R.\ 1993, \mnras, 260, 675. doi:10.1093/mnras/260.3.675
\bibitem[Kippenhahn \& Weigert(1967)]{1967ZA.....65..251K} Kippenhahn, R. \& Weigert, A.\ 1967, \zap, 65, 251
\bibitem[Fowler \& Hoyle(1964)]{1964ApJS....9..201F} Fowler, W.~A. \& Hoyle, F.\ 1964, \apjs, 9, 201. doi:10.1086/190103












\end{thebibliography}
\end{document}